\documentclass[aps,pra,twocolumn,floatfix,showpacs,superscriptaddress]{revtex4-1}

\usepackage[T1]{fontenc}
\usepackage[breaklinks, colorlinks=true, urlcolor=blue, anchorcolor=blue, citecolor=blue, filecolor=blue, linkcolor=blue, menucolor=blue, linktocpage=true, pdfproducer=medialab]{hyperref}
\usepackage{amssymb,bbm,amsmath,graphicx,times}
\usepackage{mathptmx}
\usepackage[usenames]{color}
\usepackage[dvipsnames]{xcolor}

\usepackage{braket}

\newcommand{\konrad}[1]{ {#1} }
\DeclareMathOperator{\Tr}{Tr} 
\DeclareMathOperator{\conv}{conv}
\newcommand\cZ{\mathcal{Z}}
\newcommand{\parent}[1]{\left( {#1} \right)}
\newcommand\state{w_1\otimes\ldots\otimes w_J}

\newcommand{\dd}{\text{d}}

\DeclareMathOperator{\diag}{diag}

\begin{document}

\title{Separability gap and large deviation entanglement criterion}
%
\author{Jakub Czartowski}
\affiliation{Jagiellonian University,  Marian Smoluchowski Institute of Physics, \L{}ojasiewicza 11, 30-348 Krak\'ow, Poland}
\author{Konrad Szyma{\'n}ski}
\affiliation{Jagiellonian University,  Marian Smoluchowski Institute of Physics, \L{}ojasiewicza 11, 30-348 Krak\'ow, Poland}
\author{Bart{\l}omiej Gardas}
\affiliation{Jagiellonian University,  Marian Smoluchowski Institute of Physics, \L{}ojasiewicza 11, 30-348 Krak\'ow, Poland}
\affiliation{Institute of Physics, University of Silesia, ul. Bankowa 12, 40-007 Katowice, Poland} 
\author{Yan V. Fyodorov}
\affiliation{
Department of Mathematics, King's College London, London  WC2R 2LS, UK}
\author{Karol {\.Z}yczkowski}
\affiliation{Jagiellonian University,  Marian Smoluchowski Institute of Physics, \L{}ojasiewicza 11, 30-348 Krak\'ow, Poland}
\affiliation{Center for Theoretical Physics, Polish Academy of Sciences, Al. Lotnik\'{o}w 32/46, 02-668 Warszawa, Poland}
\date{\today}
%
\begin{abstract}
\noindent
For a given Hamiltonian $H$ on a multipartite quantum system, one is interested in finding the energy $E_0$ of its ground state.
In the separability approximation, arising as a natural consequence of 
measurement in a separable basis, one looks for the minimal expectation value $\lambda_{\rm min}^{\otimes}$ of $H$ 
among all product states. For several concrete model Hamiltonians, we investigate the difference $\lambda_{\rm min}^{\otimes}-E_0$,  called separability gap,  which vanishes if the ground state has a product structure. 
In the generic case of a random Hermitian matrix of the Gaussian orthogonal ensemble, we find explicit bounds for the size of the gap which depend on the number of subsystems and hold with probability one.
This implies an effective entanglement criterion applicable for any
multipartite quantum system: If an expectation value of a typical observable among a given state is sufficiently distant 
from the average value, the state is almost surely entangled.
\end{abstract}

\maketitle

\section{Introduction.}
Describing complex many-body physical systems one 
often postulates a suitable Hamiltonian $H$ 
and tries to find its ground state energy $E_0$. 
From the mathematical perspective, one thus faces an optimization problem
when searching for the minimal expectation value among all normalized pure states $\ket{\psi}$. That is to say, $E_0(H)=\min_{\psi}   \langle \psi| H |\psi\rangle$.
In principle, if a hermitian Hamiltonian matrix $H$ is provided, 
one can diagonalize it, find its spectrum and thus easily identify the
smallest eigenvalue $E_0$. Nevertheless, if the system in question consists of $L$ interacting particles (e.g. spins), the dimension $N$ of the matrix grows exponentially, $N=2^L$, rendering this simplistic approach ineffective for $L \gg 1$. 

Although heuristic algorithms for large systems exist~\cite{Wittek18,Gardas18}, they are most likely to fail in the high-entanglement limit~\cite{Eisert_2010}.
In such cases of practical importance one applies various methods based on quantum annealing~\cite{ST06,Bo14} and can depend on an increasing number of dedicated physical annealing systems~\cite{Lanting14,fujitsu_2018,harris_phase_2018,king_observation_2018}. Relying on this approach, however, one faces a variety of difficulties and challenges~\cite{Gardas17}. There is one particular drawback that is \emph{not} readily evident. Namely, at the end of a quantum annealing, one measures the orientation of individual spins forming the system and obtains an approximation to the ground state energy related to a product state, 
$\lambda_{\rm min}^{\otimes} (H)=\min_{\psi_{\rm sep}}   \langle \psi_{\rm sep} | H |\psi_{\rm sep}\rangle$,
where the minimum is taken over all product states,
$|\psi_{\rm sep}\rangle= |\phi_1\rangle \otimes |\phi_2\rangle \otimes \cdots \otimes |\phi_L\rangle$~\footnote{Measuring a complex system in a highly entangled energy basis (which may {\it a priori} be unknown) is practically impossible.}.
Such separable states, admitting the simplest tensor network structure with bond dimension being one~\cite{orus14}, are physically associated with the mean field like approximations.

Although for a system composed out of $L\sim 10^3$ spins selecting the optimal configuration of signs out of $2^L$ possibilities is already a great achievement, in this way one cannot obtain any approximation for the ground state energy better than the minimal product value $\lambda_{\rm min}^{\otimes}(H)$. The size of the {\sl separability gap} $\Delta_{\rm sep}(H)$, defined by the difference of both minima,
\begin{equation}
\Delta_{\rm sep}(H)= \lambda_{\rm min}^{\otimes}(H)- E_0(H),
\label{gap1}
\end{equation}
depends clearly on the analyzed Hamiltonian $H$.

The aim of the present work is to investigate to
what extent this issue poses a fundamental limitation
to the near-term quantum annealing technology.
In particular, we identify Hamiltonians for which the separability 
gap~(\ref{gap1}) becomes significant.
As for those Hamiltonians there exists a systematic 
upper bound for the precision of the separable state 
approximation commonly used by noisy intermediate scale 
device~\cite{Preskill18}, quantum annealers in particular.  

Since the latter devices are far from being perfect in many aspects~\cite{GardasDeffner18}, the measurement process they perform has not been put under theoretical scrutiny. However, as the quantum technology improves, this problem becomes more and more relevant for practical applications~\cite{Gardas18NN}. 
In this paper, we show that for a generic Hamiltonian the separable state approximation leads to a significant and systematic error of the ground state energy. Our findings allow us to formulate the large deviation entanglement criterion based on a generic, \emph{macroscopic}, observable that is applicable for any multipartite quantum system. The term ``generic Hamiltonian'' refers to a typical realization of a random Hermitian matrix pertaining to the Gaussian orthogonal ensemble of a fixed dimension.

We emphasize that it is the measurement process performed by \emph{current} (and most likely also by near-term~\cite{Ozfidan19}) quantum annealers that serves as the main motivation behind our work. As far as we know, with these machines one can only measure individual spins in the computational basis. A primary example is the D-Wave $2000$Q machine where all spins are measured in the $z$-basis to reconstruct the final (classical) energy. Here we simply pin point far reaching consequences of this fact, indicating the very limit for the underlying present-day technology. 

\section{Extreme separable values and product numerical range}
To tackle the aforementioned issue we begin with basic notions and definitions concerning spectrum of quantum systems. 
The set of possible expectation values of an operator $H$
among all normalized states, $W(H)=\{z:z=\
 \langle \psi| H |\psi\rangle\}$, is called {\sl numerical range}~\cite{GR97}.
For any hermitian matrix, $H=H^{\dagger}$ of order $N$, this set forms an interval
along the real axis between the extreme eigenvalues, $W(H)=[E_0,E_{N-1}]$,
where the eigenvalues (possibly degenerated) are ordered, $E_0\le E_1\le \dots, \le E_{N-1}$. 
  
Assume now that (i) $N=M^J$ so that the Hilbert space has a tensor structure, ${\cal H}_N={\cal H}_M^{\otimes J}$, and (ii) the product states $|\psi_{\rm sep}\rangle$ are defined.
By analogy, the set of expectation values of $H$
among  normalized product states, 
 $W^{\otimes}(H)=\{z:z=\
 \langle \psi_{\rm sep} | H |\psi_{\rm sep}\rangle\}$,
is called  {\sl product numerical range}~\cite{PGMSCZ11}.
By definition it is  a subset of $W(H)$
and for a Hermitian $H$ it forms an interval
between extreme product values,
 $W^{\otimes}(H)=[ \lambda_{\rm min}^{\otimes}, \lambda_{\rm max}^{\otimes}]$.
Product numerical range found several applications in the theory of quantum information~\cite{GPMSZ10}. 
For instance, if the  minimal product value 
of a hermitian matrix $H$ of size $d^2$ is non-negative,
then $H$ represents an entanglement witness or a positive map useful for entanglement detection~\cite{HHHH}.

\subsection{Linear chain of interacting qubits.}
The model we are going to discuss first is
motivated by the idea of finding the ground state of a physical system (consisting of interaction qubits) with spin-glass quantum annealers~\cite{Lanting14}.
After the annealing cycle has been completed, just before the final measurement, the system Hamiltonian reads~\cite{Bo14}
\begin{equation}
H = - \sum_{\langle i, j\rangle \in \mathcal{E}} J_{ij} \hat \sigma^z_i \hat  \sigma^z_j - \sum_{i \in \mathcal{V}} h_i \hat  \sigma^z_i.
\label{eq:HIsing}
\end{equation}
Here, $\hat \sigma^z_i$ is the $z$-th component of the spin-$1/2$  operator (acting on a local Hilbert space ${\cal H}_2$) associated with $i$-th qubit. Input parameters $J_{ij}$, $h_i$ are defined on a graph $\mathcal{G} = (\mathcal{E}, \mathcal{V})$, specified by its edges and vertices. They 
encode the initial problem to be solved~\cite{Lanting14}. Clearly, this Hamiltonian is classical in a sense that all its terms commute. Thus, the final measurement can be carried out on individual qubits, in any order, without disturbing the system~\cite{Zurek03}.
After that, the ground state energy is easily reconstructed from
the eigenvalues that were measured. This is of great practical importance. 
However, to become general purpose computing machines~\cite{Preskill18} near-term annealers will need to include interactions between the remaining components of
the spin operator, $\sigma_i^x$, $\sigma_i^y$~\cite{Lidar07}.

General purpose computing machines are those that realize the gate model of quantum computation to which adiabatic quantum computing is equivalent (with possible polynomial overhead); cf. Ref.~\cite{Lidar07}. Although one cannot establish this equivalence with only $ZZ$ interactions, it is sufficient to add only $XX$ or $ZX$ type of interactions to the annealer Hamiltonian to demonstrate universality~\cite{Biamonte08}.

For the sake of argument, assume that the final measurement can be accomplished faithfully. Also, let the system be shielded 
from its environment for as long as it is necessary to perform computation. Even then, there exists a fundamental 
limitation on how much information can be extracted from the system by measuring it in the computational basis. 
We demonstrate this feature studying a chain of $L$ spins with a
nearest neighbour coupling -- the 1D Heisenberg model in the transverse magnetic field~\cite{Arnesen01},
\begin{equation}
H = -\sum_{i=1}^{L-1}\left(\hat \sigma_{i}^z \hat \sigma_{i+1}^z + \hat \sigma_{i}^x \hat \sigma_{i+1}^x\right)
-h\sum_{i=1}^L \hat \sigma_i^z.
\label{zx}
\end{equation}
\begin{figure}[t!]
	\includegraphics[width= \columnwidth]{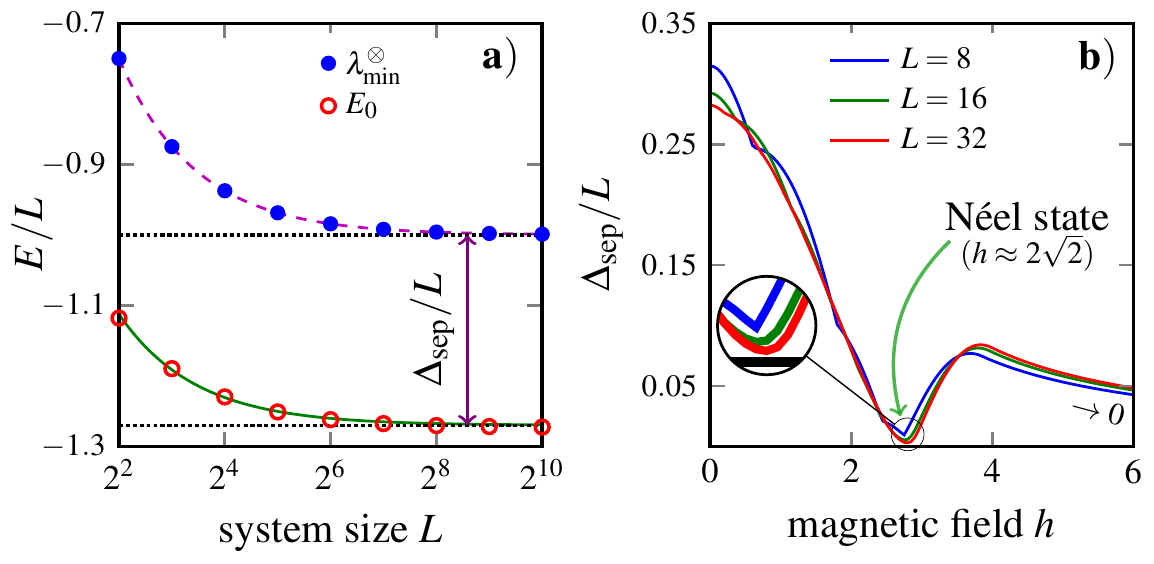}%
	\caption{A numerical solution obtained for the 1D Heisenberg model in Eq.~(\ref{zx}).
		  Panel {\bf a)} shows the ground state energy, $E_0$, together with the minimal reachable energy, $\lambda_{\text{min}}^{\otimes}$,
		 as a function of the system size $L$. Analytic calculations yield $\lambda_{\text{min}}^{\otimes}=1/L-1$ 
		  and the best fit results in $E_0/L = 0.63/L - 1.27$.		 
		 Panel {\bf b)} shows the separability gap, $\Delta_{\text{sep}}/L:=(\lambda_{\text{min}}^{\otimes}-E_0)/L$ versus  
		 the  magnetic field $h$. 
		 The apparent local minimum at $h \approx 2\sqrt{2}$ corresponds to a N{\'e}el product state.
		\label{Fig:1}
	}
\end{figure}
Although for a general Hamiltonian it is hardly possible 
to evaluate the minimal product value analytically,
it is doable in the case of vanishing magnetic field, $h=0$.

In order to simplify the matter we assume spherical coordinates $(\theta', \phi')$ on a Bloch sphere, rotated such that the main axis lies along the $y$ axis of the standard Cartesian coordinates. Under such assumption it can be shown that expectation value on a separable state $|\Psi'_{\rm sep}\rangle = \bigotimes_{i=1}^L |\psi(\theta'_i,\,\phi'_i)\rangle$ yields
\begin{equation}
    \langle \Psi'_{\rm sep}| H |\Psi'_{\rm sep} \rangle =  \sum_{i=1}^{L-1} \sin\theta_i\sin\theta_{i+1}\cos(\phi_i-\phi_{i+1}),
\end{equation}
thus the minimal product value
reads $\lambda_{\text{min}}^{\otimes}=1 - L$.

A numerical simulation (cf. Fig.~\ref{Fig:1}{\color{blue} a}) 
shows
that the separability gap $\Delta_{\text{sep}}$
plays a crucial role for any system size.
For a large number of qubits
the gap grows linearly with the system size,
$\Delta_{\text{sep}} \approx C L$ with 
$C \approx 0.27$.
In the asymptotic limit, $L\rightarrow \infty$,
the ground state energy of (\ref{zx})
was derived analytically,
$E_0/L=-4/\pi$,
for the same system with
periodic boundary conditions~\cite{Taylor_1985,Maciazek16}.
As in this limit $E_0$ does not depend on the boundary conditions,
we arrive at the explicit result for the 
asymptotic separability gap,
\begin{equation}
\Delta_{\text{sep}}(H) 
=\lambda_{\text{min}}^{\otimes} - E_0  
\xrightarrow[L \to \infty]{}
(\frac{4}{\pi}-1)L.
\label{gap2}
\end{equation}
This implies a systematic error if 
the ground state energy is approximated 
by reconstructing the ground state by an optimal product state. To put it differently, in this case the true minimal energy of the 
system can never be reached
by any annealing procedure. 

The separability gap is maximal at $h=0$ and vanishes 
in the case of very strong fields, $|h| \gg 1$,
for which the interaction part of $H$ can be neglected.
Interestingly, this dependence is not monotonic,
as the separability gap $\Delta_{\text{sep}}$
exhibits its minimum at $h \approx 2\sqrt{2}$. 
At this value of the field the gap tends to zero,
since the ground state of the system becomes separable N{\'e}el product state~\cite{Niederberger10}.

\subsection{Toy model with interaction between all subsystems}
Consider an arbitrary Hamiltonian $H$ 
describing a system of $L$ qudits and acting on the space of dimension $d^L$.
If the eigenstate $|\psi_0\rangle$ corresponding to the 
eigenvalue $E_0$ is separable, the separability gap 
vanishes by definition. However, the reverse implication does not hold,
as the gap  $\Delta_{\rm sep}$ can be arbitrary small even if two
eigenstates with the smallest energies, $E_0$ and $E_1$ 
are strongly entangled.

To investigate this problem consider a model Hamiltonian matrix
representing a two-qubit system
\begin{equation}
H_2  =
\begin{pmatrix}
             0 & 0 & 0 & 1 \\
             0 & 0 & a & 0 \\
             0 & a  & 0 & 0\\
             1 & 0  & 0 & 0\\
\end{pmatrix}
 =:A(1,a,a,1),
\label{ham2}
\end{equation}
where
$A(x_1,\dots, x_N)$
denotes a matrix 
with the vector $x$ at the antidiagonal
and zero entries elsewhere.
%
Then the Hamiltonian 
can be written as~\cite{Taylor_1985} 
\begin{equation}
%
    H_2 = \left(2 + 2a\right)\sigma_x^{\otimes 2} + \left(2a - 2\right) \sigma_y^{\otimes 2}.
\end{equation}
We shall assume that  $a\in [0,1]$,
so the ordered spectrum of $H_2$ reads $(-1,-a,a,1)$
and $E_0=-1$.
In the non-degenerate case, $a \in (0,1)$,
all the eigenvectors of $H_2$ are maximally entangled
and they form the Bell basis~\cite{nielsen_chuang_2010}.
Due to the special form of $H_2$ 
it is possible to perform optimization over product states analytically. 
By assuming angular parametrisation on the Bloch sphere $|\psi(\theta_i, \phi_i)\rangle = (\cos \theta_i/2, e^{i\phi_i}\,\sin \theta_i/2)$ for each qubit, we arrive at the expectation value of $H_2$ on a product state $| \psi_{\rm sep}(\theta_1, \theta_2, \phi_1, \phi_2) \rangle \equiv |\psi_{\rm sep}\rangle$
\begin{equation}
  \begin{split}
    \langle \psi_{\rm sep}| H_2 | \psi_{\rm sep} \rangle &= \frac{1}{2}\sin\theta_1\sin\theta_2 \\
    &\times
    \left[\cos(\phi_1+\phi_2) + a\cos(\phi_1 - \phi_2)\right], 
    \end{split}
\end{equation}
which is to be minimized. 
By setting $\theta_1 = \theta_2 = \pi/2,\,\phi_1 + \phi_2 = \pi$ and $\phi_1 - \phi_2 = \pi$ 
we arrive at the minimal value
$\lambda_{\rm min}^{\otimes}(H_2)=-(1+a)/2$.
Note that the separability gap
$\Delta_{\rm sep}=(1-a)/2$ is the largest for $a=0$
and it vanishes for $a=1$.

Analyzing dimension of a subspace which contains at
least a single separable state 
one can show \cite{PGMSCZ11} that for a hermitian matrix of order $N=4$ 
the minimal product value is not larger than the energy of the first excited  state,
$E_0 \le \lambda_{\rm min}^{\otimes} \le E_1$,
so in this case the separability gap
is bounded, $\Delta_{\rm sep} \le \Delta_1=E_1-E_0$.
Hence in the limit $a \to 1$ the spectrum of $H_2$
becomes degenerated and thus the separability gap vanishes.

Let us now generalize the above model for $L$ qubits
by considering a symmetric, antidiagonal real matrix of size $N=2^L$ such that $(H_L)_{1,N} = (H_L)_{N,1}= 1$
and all other entries equal to zero. This Hamiltonian captures an all-to-all type of interactions between qubits and can be written in a compact form,
$H_L=\sigma_+^{\otimes L} + \sigma_-^{\otimes L}$,
where $\sigma_\pm = \sigma_x \pm i \sigma_y$.
The only non-zero eigenvalues are $\pm 1$ and thus $E_{0} = -1$. To calculate the minimum value over the product states $\lambda^\otimes_\text{min}$ 
we again resort to the polar coordinates on the Bloch ball and define state $|\Psi_{\rm sep}\rangle = \bigotimes_{i=1}^L |\psi(\theta_i,\phi_i)\rangle$. Calculating the expectation value on such state yields
\begin{equation}
    \langle \Psi_{\rm sep} | H_L
    | \Psi_{\rm sep} \rangle = 2^{1-L} \left(\prod_i^L \sin(\theta_i)\right) \cos\left(\sum_{j=1}^L\phi_j\right),
\end{equation}
which is easily minimized with $\theta_i = 0$ and $\sum_{j=1}^n \phi_j = \pi$. The resulting minimal separable expectation value, $\lambda_{\text{min}}^\otimes (H_L)=2^{1-L}$, tends to zero as $L\rightarrow \infty$ (recall that $E_{0}=-1$). 
Similar conclusions can be drawn by analyzing a family of real symmetric and  antidiagonal Hamiltonians with no more than $2L$ non-zero entries,
\begin{equation}
\label{eq:fam}
(H'_L)_{i,j} = 
\begin{cases} 
      a_k & k = 0,\hdots,L-1,\\
      & (i,j)=(1+k,N-k) \vee (i,j)=(N-k,1+k) \\
      0 & \text{otherwise}.\nonumber
\end{cases}
\end{equation}
In particular, setting $a_1=1$ one obtains $E_0=-1$, thus the support of the spectrum is $[-1,1]$. On the other hand, one can show using analogous method as before that
\begin{equation}
\lambda_{\rm min}^\otimes(H'_L) = 2^{1-L} \sum_{k=1}^L |a_k|.
\end{equation}
Hence, the above model extends the family of Hamiltonians for which $\lambda_{\rm min}^\otimes$ tends to zero in the case of a large number of qubits,
despite the support of $H_L$ being fixed.

As we will shortly see, this non--intuitive property
is characteristic for generic Hamiltonians. This is an important result especially since $\lambda_{\rm min}^{\otimes}(H)$ can \emph{not} be calculated analytically in general~\cite{SZ09} and furthermore all known numerical methods are restricted to small system sizes (cf. Appendix I).

\section{Generic Hamiltonians of $L$-qubit systems}
\begin{figure}[t!]
	\includegraphics[width=1 \columnwidth]{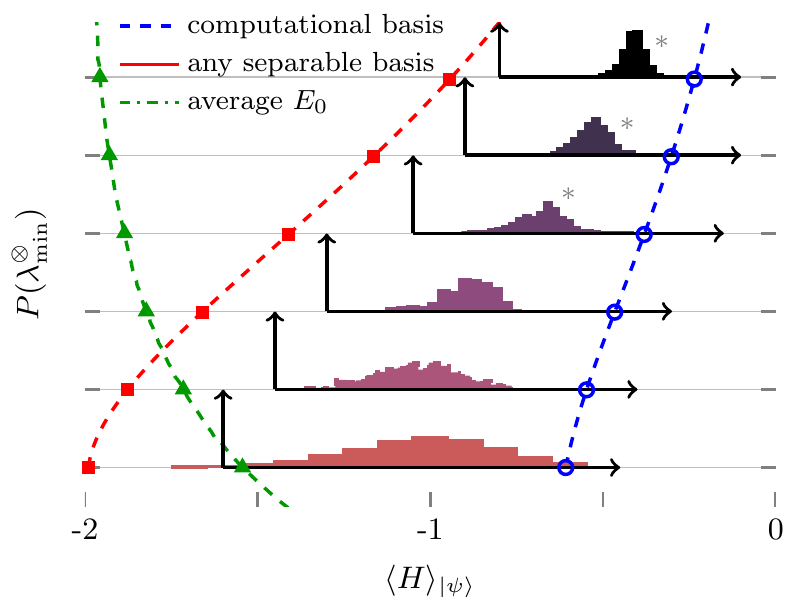}%
	\caption{Collection of six distributions
	  $P(\lambda^\otimes_{\text{min}})$
	  of   minimal separable expectation values
	   for generic GOE Hamiltonians of dimension
	     $N=2^L$ for $L=3,\dots , 8.$
	     Red squares (blue dots) denote 
	     asymptotic lower (upper)  bounds for $\lambda^\otimes_{\text{min}}$
	obtained in Eq. \eqref{bounds} and (B15) (in Appendix) and with fixed $M=4$,
	green triangles represent the average ground state energy $E_0$. Dashed lines are plotted to guide the eye.
	}
	\label{fig:GOE}
\end{figure}
The situation in which separable states do not approximate well the ground state is in some sense generic (or typical). To substantiate this statement let us consider random hermitian matrices drawn from the Gaussian Orthogonal Ensemble (GOE) of size $N=2^L$, which describe Hamiltonians acting on $L$ qubits. For each sample matrix $H$ we wish to determine minimal eigenvalue $E_{0}$ and estimate minimal separable expectation value $\lambda_{\text{min}}^\otimes$. 
Due to the concentration of measure in the limit of a large system size these quantities become self-averaging, so that for a typical realization their values are close to the ensemble averages~\cite{RM18}.

Generically no product states are found in subspaces with dimension comparable to $N$. In the case of $L$ qubits a subspace of dimension $2^L-L-1$ almost surely (a.s.) contains no product state~\cite{walgate}. It is therefore reasonable to expect that the range of expectation values of a GOE Hamiltonian over product states shrinks with increasing system size: product states are superpositions of \emph{almost all} eigenstates of the Hamiltonian. This behavior holds true as it is a consequence of the following two results:

{\bf Proposition 1.} {\sl Consider a generic Hamiltonian 
represented by a GOE matrix $H$ of size { $N=M^J$,} with $M,J\gg 1$
normalized as $\langle {\rm Tr}H^2\rangle=N$,
so that the minimal energy asymptotically reads 
{$E_{0}\rightarrow -2$.}
Then the minimal value $\lambda^{\otimes}_{\rm min}$
among all product states of the $J$--partite system
satisfies the following estimates with probability one  (almost surely),}

\begin{equation}
\label{bounds}
-\frac{2 J}{\sqrt{N}}
\
\underset{\rm a.s.}{\le}
\
\lambda^{\otimes}_{\rm min}
\
\underset{\rm a.s.}{\le}  
\
-\sqrt{\frac{4 \ln N }{N}}.
\end{equation}

{\bf Proposition 2.}. {\sl The above estimates work also for the 
partition of total space into $L$ qubits. Let us assume that $M=2^K$ so that $N=2^L$ with $L=K+J$, and any state separable with respect to partition ${\cal H}_2^{\otimes L}$ is separable for splitting ${\cal H}_M^{\otimes J}$ as well.}

To derive the upper estimate note that the diagonal entries of $H$
correspond to expectation values among product states 
$|i_1 i_2 \dots i_J\rangle$. 
For any random  GOE matrix of size $N$ its diagonal,
$D=\diag H$, is a sequence of $N$ numbers independently drawn from the normal distribution ${\cal N}(0,\sqrt{1/N})$. Therefore, the typical minimal entry on diagonal $\langle \min~D\rangle_{\rm GOE}$ behaves as $-\sqrt{4\ln N/N}$~\cite{leadbetter2012extremes} and leads to the right inequality in~(\ref{bounds}).
The reasoning leading to the lower estimate
relies partly upon the use of the so-called ``replica trick'' and saddle point approximation (cf. \cite{Fyodorov02} and 
Appendix II for a more detailed analysis).

Fig.~\ref{fig:GOE} presents histograms of  the smallest
separable expectation value $\lambda^\otimes_{\text{min}}$ 
obtained for a sample of $10^3$ random Hamiltonians 
from the Gaussian Orthogonal Ensemble 
of size $N=2^L$. Numerical data 
are obtained by the algorithm described in Appendix I
or a standard optimization algorithm ($*$).
Results obtained confer to the bounds
(\ref{bounds}).
	The lower bound corresponds to a measurement of the energy in an \emph{optimized} separable basis, while the upper one	to a measurement carried out in the \emph{fixed} separable basis.

Proposition 1 implies that for a typical random matrix $H$
 acting on an $L$--qubit system
 $\lambda_{\text{min}}^\otimes(H) \rightarrow 0$ 
 with probability one, 
although $E_{0}(H)\rightarrow -2$. 
This observation implies that for a large system described by a generic Hamiltonian the separability gap is constant,
$\Delta_{\rm sep} \to 2$,
so it is not possible to obtain any accurate estimation of the ground state energy, if the measurement is performed in {\sl any} separable basis.

It is worth emphasizing that the above observation has key consequences for the theory of multipartite entanglement in large quantum systems: Measuring any generic observable $A$ of a composed system of  total dimension $N$  
in a separable state yields outcome close to \konrad{the average of eigenvalues} 
$\bar A= {\rm Tr}A/N$. 
This above statement can be connected with earlier results
of Wie{\'s}niak et al.~\cite{WVB05}, who proposed to consider macroscopic quantities, like magnetic susceptibility,
as entanglement witness. In fact our observation can be formulated in a similar spirit.

Any generic hermitian observable $A$ of order $N=M^J$
allows one to construct two dual entanglement witnesses,
corresponding to both wings of the semicircular spectrum,
$W_{\pm }(A):={\mathbbm I} \pm c_\pm A$, 
such that any negative expectation value, 
Tr$\rho W_{\pm} < 0$, implies entanglement of the state $\rho$.
The actual value of the parameter, 
$c_{\pm}=N/( 
J \sqrt{\Tr A^2} \mp \Tr A)$,   
as a function of the total system size $N$,
number of parties $J$, 
mean value and the variance of $A$, follows from the bound~(\ref{bounds}),
since it implies that the matrix $W_{\pm }$ is positive among all states separable
with respect to the partition
${\cal H}_N={\cal H}_M^{\otimes J}$.
The above result can be reformulated into  
the following simple, yet a very general
{\sl large deviation entanglement criterion}.

\begin{figure}[t!]
	\includegraphics[width= \columnwidth]{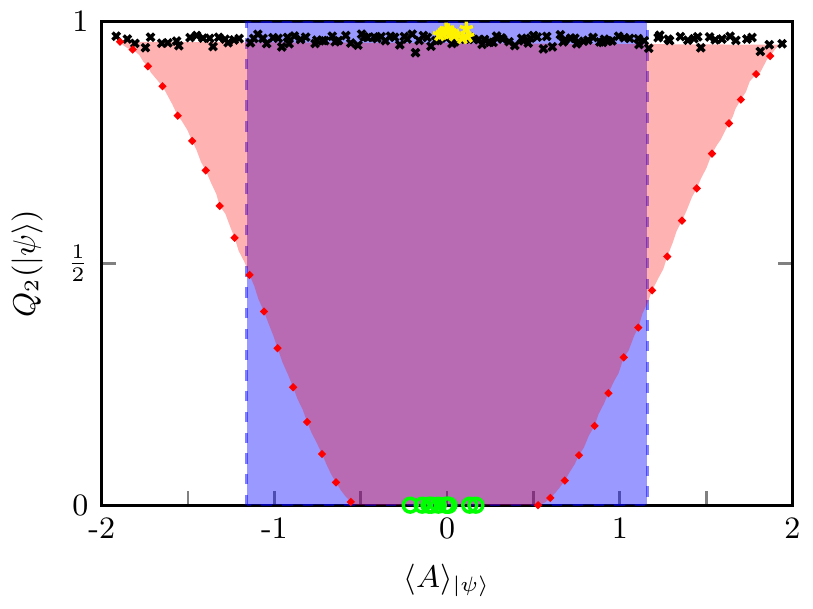}%
	\caption{Range of allowed values for pure states of the system consisting of $L=7$ qubits
	in the plane spanned by the
	   expectation value $\langle A\rangle_{|\psi\rangle}$ of a  GOE observable $A=H$ of size $N=2^L$,
	    and the Meyer-Wallach measure of 
	entanglement defined in Eq.~(\ref{mw}). Black crosses denote eigenstates of $H$, red region -- numerically determined range attained by pure states, shaded blue region denotes the bound $|{\rm Tr}A\rho - \bar A| \le 2 L \sqrt{\Tr A^2}/N^2$ implied by
	Eq.~(\ref{ldec}),
	 beyond which the states are entangled. In addition, yellow circles represent a sample of $10$ random pure states, green squares -- $10$ random product states.
		\label{Fig:wallach}
	}
\end{figure}
Namely, if an expectation value of a typical observable
$A$ of order $N=M^J$ in the state $\rho$
is sufficiently distant from the 
barycenter of the spectrum,
$\bar A=\text{Tr} A/N$, that is when
\begin{equation}
\label{ldec}
|{\rm Tr}A\rho - \bar A| \ > \ 
2 J \sqrt{\Tr A^2}/N^2,
\end{equation}
then the state $\rho$ is almost surely entangled
with respect to the partition
into $J$ subsystems with $M$ levels each.
                        
Hence this criterion belongs to the class
of double--sided entanglement witnesses $2.0$
recently analyzed in~\cite{BCH18}.
Note that the reasoning holds in one direction only
as there exist also entangled states 
for which the expectation value is close to the mean $\bar A$.
However, numerical computations confirm a natural 
conjecture that the larger the absolute value of the deviation,
$\delta=|\langle \phi|A|\phi\rangle -\bar A|$,
the larger average entanglement of the analyzed state $|\phi\rangle$ 
(cf. numerical results presented in Fig.~\ref{Fig:wallach}). 
To quantify entanglement of pure states of an $L$-qubit system
we used the family of measures introduced by Meyer and 
Wallach~\cite{meyer2002global}, which are based on the
linear entropy of reduced states 
averaged over all possible reductions consisting of $k$ subsystems,
\begin{equation}
    \label{mw}
    Q_k(|\psi\rangle) = \frac {2^k} {2^k -1} {L \choose k}^{-1}\sum_{X: |X|=k} S_{\text{lin}} (\rho_X) ,
\end{equation}
where $S_{\text{lin}} (\rho) = 1-\text{Tr} \rho^2$ being the linear entropy of a state $\rho$ of dimension $2^k$.
This function captures the mean entanglement of  $k$-qubit subsystems
with the rest of the system.
Although Fig.~\ref{Fig:wallach} depicts data obtained for $Q_2$,
similar results were also analyzed for other measures of entanglement, including quantities $Q_k$  with $k=1,\dots, L$.
All these results support the statement that the deviation 
of the expectation value $\langle A\rangle_{\psi}$ beyond
the bounds~(\ref{ldec})
can be used to quantify the degree of entanglement 
of the analyzed state $|\psi\rangle$.

For comparison Fig.~\ref{Fig:wallach} 
contains also data for random separable states and generic random states,
which are known to be highly entangled
\cite{ZS01,HLW06}. The set of 
separable pure states has
a lower dimension and carries zero measure in the entire set of all pure states,
so its projection $W^{\otimes}(A)$
onto an axis
determined by the observable $A$
is typically much smaller
than the entire range $W(A)$.
Asymptotically, in the
limit of large dimension $N$
of the Hilbert space, the ratio of the
volumes of both sets tends to zero.

\section{Discussion and outlook}
In this work we have investigated to what extent the near-term quantum annealing technology may become fundamentally limited by its intrinsic measurement process allowing to ask only \emph{yes} or \emph{no} questions to individual qubits. This type of ``polling''
on a quantum system is probably the most natural one and definitely the easiest to realize experimentally. Unfortunately, as we have argued, it does \emph{not} allow to extract all relevant information from the system in question.

In particular, we analyzed the separability gap and showed that it is
non-zero for several model Hamiltonians acting on multipartite quantum systems.
Moreover, we studied  Hamiltonians constructed by random matrices from
the Gaussian orthogonal ensemble and demonstrated that for such a generic
Hamiltonian involving L qubits the minimal value of energy
$\lambda_{\rm min}^\otimes$ among all product states is significantly larger than the
ground state energy $E_0$.
Thus making use of near-term quantum annealers, 
in which the final result is 
obtained by independent measurements 
of each of $L$ qubits
and corresponds to a product state, can \emph{not}
provide a reliable approximation for the ground state energy of a typical problem.  
Furthermore, we formulated an entanglement criterion
based on the expectation value 
of a generic observable $A$ 
among an arbitrary state $\rho$ of a composed 
quantum system and showed that ${\rm Tr}\rho A$ provides direct information concerning the degree of entanglement of the investigated state $\rho$.

\section*{Acknowledgements}

It is a pleasure to thank Jacek Dziarmaga, 
B{\'a}lint Koczor,  Tomasz Maci{\k a}{\.z}ek, Andy Mason, Zbigniew Pucha{\l}a, Marek Rams, Ilya Spitkovsky and Jakub Zakrzewski
for inspiring discussions. Financial support 
by Narodowe Centrum Nauki under the grant number 
2015/18/A/ST2/00274 (K{\.Z})
and 2016/20/S/ST2/00152 (BG) is gratefully acknowledged. The research at KCL was supported by EPSRC grant EP/N009436/1. This research was supported in part by PL-Grid Infrastructure.

\section{Appendix}
\setcounter{equation}{0}
\setcounter{figure}{0}
\renewcommand{\theequation}{A\arabic{equation}}
\renewcommand{\thefigure}{A\arabic{figure}}
\subsection{Numerical technique to estimate  
$\lambda^\otimes_{\text{min}}(H)$}
\label{app:A}
We briefly sketch here the approach employed in this work
to calculate the separability gap for a random Hamiltonian
pertaining to Gaussian Orthogonal Ensemble (GOE), 
in the case of a small system size (up to $N=2^8$). The ground state energy $E_0$ can be obtained easily in this case. The algorithm used for calculation of  minimal separable expectation -- $\lambda^\otimes_{\text{min}}$, on the other hand, utilizes the divide and conquer strategy~\cite{knuth98}. 
To begin with, let us consider a general case of minimizing expectation value of $\langle \alpha\otimes \beta| H |\alpha\otimes\beta\rangle$, where $\ket{\alpha}\in{\cal H}_2$
is a qubit state and $\ket{\beta}$ 
 belongs to a $d$-dimensional space ${\cal H}_d$.
 The expectation value can be rewritten as
\begin{equation}
    \langle \alpha\otimes \beta| H |\alpha\otimes\beta\rangle = \bra{\alpha}H_{\ket{\beta}}\ket{\alpha}, 
\end{equation}
where $H_{\ket{\beta}}=\Tr_B [H (1 \otimes \ket{\beta}\bra{\beta})]$ is a matrix of size 2. If $\ket{\beta}$ is fixed, further optimization over $\ket{\alpha}$ is trivial: the result is minimal eigenvalue of $2\times2$ hermitian matrix $H_{\ket{\beta}}$:
\begin{equation}
    \min_{\ket{\alpha}} \bra{\alpha}H_{\ket{\beta}}\ket{\alpha} = \frac{\Tr H_{\ket{\beta}}}{2}-\sqrt{\left(\frac{H_{\ket{\beta}}}{2}\right)^2 - \det H_{\ket{\beta}}}.
\end{equation}
The above expression can be written in a more succinct form. Let $H_i=\Tr_A[ H( \sigma_i \otimes 1) ]$ (with $\sigma_0=1_2$). 
Then, the above expression becomes
\begin{equation}
    \frac{\langle H_0 \rangle}{2} - \frac12 \sqrt{\langle H_1\rangle^2 +\langle H_2\rangle^2 +\langle H_3\rangle^2 },
\end{equation}
where all averages are taken over the $d$-dimensional vector $\ket{\beta}$. The minimization of $\langle \alpha\otimes \beta| H |\alpha\otimes\beta\rangle$ can be now interpreted as minimization of \emph{convex} function over $4$-dimensional \emph{convex} set of simultaneous expectation values called numerical range:
\begin{equation}
\begin{split}
W(H_0,H_1,H_2,H_3) = \conv \{ (\langle& H_0\rangle, \langle H_1\rangle, \langle H_2\rangle,\langle H_3\rangle)_{\ket{\beta}} :\\
&\ket{\beta}\in{\cal H}_d\}.
\end{split}
\end{equation}
This problem is easily numerically solved with an arbitrary high accuracy.

Solution of the ${\cal H}_2 \otimes {\cal H}_d$ case can be leveraged to the more general  ${\cal H}_2^{(\otimes k)} \otimes {\cal H}_d$, where $k\in \mathbb{N}$: using the procedure described above, it is possible to determine arbitrarily close approximation of the set 
\begin{equation}
\begin{split}
W^\otimes (H_0,H_1,H_2,H_3) = \conv\{& (\langle H_0\rangle, \langle H_1\rangle, \langle H_2\rangle,\langle H_3\rangle)_{\ket{\gamma}} \\
&: \ket{\gamma}\in{\cal H}_2\otimes{\cal H}_d, \ket{\gamma}=\ket{\alpha\otimes \beta}\}.
\end{split}
\end{equation}
This (convex) set $W^\otimes$ can then be used in place of $W$ in calculation of $\lambda^\otimes_{\text{min}}$ -- the result is minimal energy over separable states in tripartite case. This result can be alike used further -- the recursive structure provides natural extensions. Complexity of the algorithm is exponential, owing to the NP-completeness of the problem, but it is possible to determine certified lower and upper bounds of $\lambda_{\text{min}}^\otimes$ this way in deterministic time and linear space complexity.

\setcounter{equation}{0}
\setcounter{figure}{0}
\renewcommand{\theequation}{B\arabic{equation}}
\renewcommand{\thefigure}{B\arabic{figure}}

\subsection{Lower estimate in Eq.~(\ref{bounds}) of main text}
\label{app:B}
In this section we provide a reasoning, which leads to the left hand side of Eq.~(\ref{bounds}) in the main text for any generic Hamiltonian $H$ of order $N=M^J$,
where $M$ denotes a high ($M\gg 1$), but otherwise arbitrary dimension of each subsystem, while  $J$ stands for their number.
The method used relies upon the use of the 
so-called ``replica trick'', which is a powerful but
not fully rigorous method of theoretical physics. 

We wish to show that the minimal separable expectation value, $\lambda_{\rm min}^{\otimes}(H)$, vanishes in the large system limit, $N\rightarrow\infty$. Here we 
analyze separability with respect to partition of the system
into $J$ subsystems of size $M$ each. 
Due to effect of concentration of measure the above quantity
is "self-averaging",
which means that its distribution becomes strongly localized around the expectation value. 
Therefore it is sufficient to \konrad{study the average value and} demonstrate that
\begin{equation}
\label{tmv}
\langle \lambda_{\text{min}}^\otimes \rangle \sim - 2J/\sqrt{N},    
\end{equation}
where the brackets denote the ensemble average, and
\begin{equation}
\lambda_{\text{min}}^\otimes=\min_{\ket{w_\otimes}} \bra{w_\otimes} H \ket{w_\otimes}, 
\end{equation}
with $w_\otimes :=\state$~\cite{Fyodorov14}. 
To this end, we assume a hermitian Hamiltonian $H$ drawn from GOE with scale parameter $a$ such that $\langle \Tr H^2\rangle =a N$. 

To begin with, we introduce the partition function~\cite{huang1987}
\begin{equation}
\label{eq:statsum}
\cZ_\beta=\int \exp \parent{-\beta \braket{w_\otimes | H | w_\otimes}}{\dd w_\otimes}, 
\end{equation}
where $\beta$ plays a role of the inverse temperature. Here, $\dd w_i$ denotes the integration measure over a single qubit space. Then, the typical separable expectation value, Eq.~(\ref{tmv}), can be found
as the zero-temperature limit of the associated free energy:
\begin{equation}
\label{limit}
\langle \lambda_{\text{min}}^\otimes \rangle
= -\lim_{\beta\to\infty} \frac{\ln\cZ_\beta}{\beta}.
\end{equation}
To calculate the latter limit, consider the following function defined for positive integer $n$,
\begin{equation}
\label{pos}
\cZ_\beta^n = \int \exp \parent{-\beta \sum_{i,j=1}^n \braket{w_\otimes^{(i)} | H | w_\otimes^{(i)} }} \prod_{i=1}^n \dd w_\otimes^{(i)},
\end{equation}
which is the $n$-th power of $\cZ_\beta$. Then we can formally write
\begin{equation}
\label{eq:avms2}
 \langle \ln \cZ_\beta \rangle = \left.\frac{\dd}{\dd n} \left\langle\cZ_\beta^n\right\rangle\right|_{n=0},
\end{equation}
which is interpreted as a derivative of an analytic continuation of $\left\langle \cZ_\beta^n\right\rangle$. 
This average can be further simplified using the equality which holds for any matrix $X$
\begin{equation}
\label{sim}
\braket{\exp(-\beta \Tr H X)} = \exp[a \beta^2 \parent{\Tr X_H^2}/2]
\end{equation}
with $a$ being the scaling parameter of the GOE. Here, $X_H$ denotes a hermitian part of a
matrix $X$.
Therefore,
\begin{equation}
\langle \cZ_\beta^n \rangle = \int \exp \parent{\frac12a \beta^2 \sum_{i,j=1}^n \braket{w_\otimes^{(i)} | w_\otimes^{(j)} }^2 } \prod_{i=1}^n \dd w_\otimes^{(i)}.
\end{equation}
By introduction of a collection of matrices $Q^{(i,j)}_k=\braket{u_k^{(i)}| u_k^{(j)}}$, so that
\begin{equation}
\label{eq:change}
\prod_{i=1}^n \dd u_\otimes^{(i)} = C(n,M)^J \prod_{i,j=1,\ldots n}^{k=1\ldots J} \dd Q^{(i,j)}_k \parent{\prod_{k=1}^J \det Q_k}^{(M-n-1)/2},
\end{equation}
where the number $C(n,M)$ does not depend on $J$~\cite{Fyodorov07}, the above integral can be written in a form
suitable for the saddle point approximation~\cite{Fyodorov02,Ronchetti1997}.
The domain of integration over matrices $Q$ goes
over positive-definite matrices of size $n$
with diagonal entries fixed to be unity.
Making use of this approximation one arrives at
an expression,
\begin{eqnarray}
\label{eq:aq}
\braket{\cZ_\beta^n} &=& \int\exp 
\left\{ 
\frac{M}2 
\overbrace{
\left[
\beta^2 \sum \parent{Q^{(i,j)}_1\ldots Q^{(i,j)}_J}^2 + \sum \ln \det Q_k
\right]
}^{\Phi}
\right\} 
\nonumber \\ 
 & \times&   C(n,M)^J \prod  \dd Q^{(i,j)}_k \parent{\prod \det Q_k}^{-(n+1)/2}.
\end{eqnarray}
where we chose a nonstandard GOE scaling, $a=M=N^{1/J}$.

In the limit of large $M$, this integral is dominated by the maximum of the exponent argument. Furthermore, $C(0,M)=1$ and 
\begin{equation}
\label{eq:sadle}
\braket{\cZ_\beta^n} \sim \lim_{n\rightarrow 0} \frac{ M \Phi(Q_{\text{optim}})}{n}.
\end{equation}
Henceforward, we assume that $Q_1=\cdots Q_J=Q$, where $Q$
is parametrized with a single parameter $q$:
\begin{equation}
\label{eq:Q}
Q(q)=\begin{pmatrix}1&q&\ldots&q\\q&1&\ldots&q\\\vdots&\vdots&\ddots&\vdots\\q&q&\ldots&1\end{pmatrix}.
\end{equation}
Such an ansatz is compatible with
properties of maximal $\Phi[Q(q)]$ 
at low temperature (high $\beta$).

By calculating maximum of $\Phi[Q(q)]$ one can easily determine that $q=1-1/\beta$ in the limit $\beta\rightarrow\infty$. 
Thus,
\begin{equation}
\begin{split}
\Phi[Q(q)]&=\sum \parent{Q^{\circ J}}^2_{i,j}+\sum \ln\det Q_k\\
&=\frac{n(n-1)}{2} q^{2J} \\ 
&+J\{(n-1)\ln [1-q]+\ln[1+q(n-1)\},
\end{split}
\end{equation}
where $\circ$ denotes the Hadamard (i.e. elementwise) product of matrices. Therefore the following estimate holds
\begin{equation}
\label{eq:sadle3}
\begin{split}
\braket{\cZ_\beta^n} &\sim \frac{M}{2} \parent{\frac{J q}{1-q} - \frac{q^{2J}}{2}+J \ln [1-q]}\\
&=\left[J(\beta-1) + \frac{(1-\beta^{-1})^{2J}}{2} - J \ln \beta\right].
\end{split}
\end{equation}
Finally, taking the limit $\beta\to\infty$ in Eq.~(\ref{limit}) we obtain $\langle \lambda_{\text{min}}^\otimes \rangle \sim -MJ$, with $M=N^{1/J}$, where $N$ is the total system size, $J$ is the number of partitions and $M$ is their local dimension. Since we have worked with the scaling $a=M$, the ensemble average, $\langle \lambda_{\text{min}}^\otimes \rangle$, needs to be compared to the average minimal eigenvalue, $E_0$. Then
we arrive at the desired expression 
\begin{equation}
\label{eq:appres}
\frac{\langle \lambda_{\text{min}}^\otimes \rangle}{E_0} =  \frac{-J N^{1/J}}{-N^{(1+J^{-1})/2}}=J N^{-(1-J^{-1})/2}.
\end{equation}
\konrad{We have assumed that $M\gg 1$ such that the saddle point method can be used. Let us now consider the case of $N\rightarrow \infty$; $M\gg 1$ is kept constant and $L\rightarrow\infty$. In this limit the following holds}
\begin{equation}
\label{B16}
\frac{\langle \lambda_{\text{min}}^\otimes \rangle}{E_0} = \frac{\log_M N}{\sqrt{N}}.
\end{equation}
This demonstrates that the estimate~(\ref{tmv}) holds for $J\gg 1$,
what completes the reasoning concerning Proposition 1.

Therefore, when dimension $N=M^J$ increases, the minimal separable expectation value $\lambda_{\rm min}^{\otimes}$ of generic Hamiltonian of size $N$ with respect to partition ${\cal H}_M^{\otimes J}$ approaches 0.

Let us now proceed to Proposition 2.
Formally, to conduct the proof we 
require that $M=N^{1/J} \gg 1$.
Let us now assume that the local dimension 
forms a power of two, $M=2^K$, so the total
dimension reads $N=M^J=2^L$ with $L=K+J$.
Any state $|\psi\rangle$ entangled with respect to the
partition of the entire system into $J$ subsystems of size $M$
is also entangled with respect to the finer partition into 
$L$ qubits. 
Therefore the estimate~(\ref{B16})
holds also for the physically motivated partition
${\cal H}_N={\cal H}_2^{\otimes L}$
and implies  that the ratio
$\langle \lambda_{\text{min}}^\otimes \rangle / E_0$ 
tends to zero in the limit $N\to \infty$.


\bibliographystyle{apsrev4-1_nature}
\bibliography{bib} 
\end{document}